\documentclass[prb,aps,amsmath,amssymb,showpacs,floatfix,twocolumn]{revtex4}
\usepackage{graphicx,epsfig}
\usepackage[usenames]{color}
\newcommand{\be}{\begin{equation}}
\newcommand{\ee}{\end{equation}}
\newcommand{\bea}{\begin{eqnarray}}
\newcommand{\eea}{\end{eqnarray}}
\newcommand{\ul}{\underline}

\begin{document}
\title{Fluctuation-exchange approximation theory of the
non-equilibrium singlet-triplet transition}   
%\bigskip
\author{B. Horv\'ath$^1$, B Lazarovits$^1$, and G. Zar\'and$^{1,2}$}
\affiliation{
$^1$Theoretical Physics Department, Institute of Physics, Budapest University of 
Technology and Economics, Budafoki \'ut 8, H-1521 Hungary\\
$^2$Freie Universit\"at Berlin, Fachbereich Physik, Arnimallee 14, D-14195 Berlin, Germany}

\begin{abstract}
As a continuation of a previous work [B. Horv\'ath et al., 
Phys. Rev. B {\bf 82}, 165129 (2010)], here we extend the so-called 
Fluctuation Exchange Approximation (FLEX) to study 
the non-equilibrium singlet-triplet transition. We show that, while 
being relatively fast and a conserving approximation, FLEX is able to
recover all  important features of the transition, including the
evolution of the  linear conductance throughout the transition, 
the two-stage  Kondo effect on
the triplet side,  and the gradual opening of the singlet-triplet gap
on the triplet  side of the transition. 
A comparison with numerical renormalization group calculations also shows that FLEX
captures rather well the width of the Kondo resonance.
FLEX thus offers a viable route to describe correlated multi-level 
systems under non-equilibrium conditions, 
and, in its rather general form, as formulated here, it 
could find a broad application in molecular electronics 
calculations. 
\end{abstract}

\maketitle

\section{Introduction}

In the past decades, fast and surprising development has taken 
place in the field of molecular electronics. Experimentalists 
succeeded in contacting and gating a variety of molecules 
\cite{heersche,natelson,roch,heersche:other,park722,park725}, 
and  gained  more and more control over them. 
They also managed to fabricate
%fabrication of 
"artificial atoms" and molecules  from 
quantum dots, 
%allowed one 
to isolate single electrons on them
and manipulate their spin \cite{vanderSypenNazarov,Marcus,SingleShot}. 

At the same time, theory seems to be legging behind, 
and describing correlated atomic and mesoscopic structures 
under non-equilibrium conditions continues to be a challenge 
for present-day theoretical solid state physics. Tremendous effort has
been devoted to the development of theoretical    
tools to  capture appropriately the transport properties 
and dynamics of these  systems,\cite{MC1,MC2,BA1,BA3,PaaskeST,PRG2,PRG3,PRG4,NRG1,NRG2,noneqDMFT,pathint}
however, with little success. 
Most methods are uncontrolled or work only for rather 
special models. Under these conditions, 
perturbative methods can be of great value: Although they are
restricted to the regime of weak interactions, they provide 
precious theoretical benchmarks for more sophisticated though less
controlled approximations. Furthermore, many experiments are 
carried out in a regime accessible by perturbation theory.

Theorists typically use the simplest possible models such as the 
(single level) Anderson model or the Kondo model to describe
correlated behavior in these systems. For these simple models 
it is well-known that perturbative approaches can work rather 
well in the appropriate parameter range. In particular, 
perturbation theory in the interaction strength $U$ of the Anderson
model is known to reproduce the generic structure of the spectral
functions,\cite{yy1,yy2,zlatic1,zlatic3} although the 
value of Kondo temperature is known to be incorrect.\cite{Hewson} Atoms and 
experimental systems are, however, far more complicated
than the single level Anderson model.\cite{Nozieres,RoschFepaper???}
Typically, magnetic impurities contain many
electrons on their $d$ or $f$ shells, and the orbital structure of
these states and the hybridization matrix elements 
as well as the Hund's rule coupling influence quantitatively 
the corresponding magnetic and physical properties. 
It would thus be important to understand the limitations of 
perturbative non-equilibrium approaches in multi-orbital systems. 
Quantum dots, where orbital structure can become important under
certain conditions, offer ideal test grounds in this regard. 
A particular and interesting example is provided
by the so-called singlet-triplet (ST)
transition.\cite{pusgla,Pust,Zarand} There  
the occupation of two nearby levels (and thereby the  spin)
of a quantum dot  with an even number of electrons
 changes due to the presence of Hund's rule
coupling. This transition has been observed in a number of different
systems such as vertical  \cite{sasaki} and lateral quantum  
dots,\cite{Wiel,Granger} 
carbon nanotubes,\cite{nanotube} or $C_{60}$ molecules.\cite{roch} A lot of
theoretical effort has also been devoted to this transition. 
In equilibrium, the transition can be understood using numerical
renormalization group methods.\cite{HofstetterVojta,Zarand} 
However, our understanding of the non-equilibrium 
situation is rather poor: 
 the  regime far from the transition could be described 
through a functional renormalization group (RG)
approach,\cite{PaaskeST}
  which is, however, not 
appropriate to describe the small bias limit on the triplet side. 
 A slave boson approach has also been applied relatively successfully
 to describe the somewhat special underscreened case, but this approach 
is rather uncontrolled and is limited to certain models.\cite{aligiasb}   

In a previous publication,\cite{prev}
 we studied the ST transition 
using a simple, perturbative approach, and showed that 
this approach works surprisingly well: It is able to capture the
physics on both sides of the transition,  i.e., the two-stage Kondo
effect on the triplet side\cite{Pust} as well as the 
local singlet formation on the singlet side, and the formation of the 
corresponding dips in the non-equilibrium  differential conductance,
${\rm d}I/{\rm d}V$.  The simple perturbative approach is, however,
not conserving in general,\cite{kb1,kb2} and furthermore, as mentioned above,
it fails to reproduce the Kondo temperature.\cite{Hewson}
Therefore, in the present work, which should be considered as an
extension of our previous study, Ref.~\onlinecite{prev}, we 
go beyond simple perturbation theory, and study, whether 
the simplest non-trivial conserving approximation, the so-called 
fluctuation exchange approximation (FLEX) is able to capture the
ST transition. This method
 has been extensively applied in
connection to high-temperature superconductivity,\cite{FLEX_b1,FLEX_b3} 
and as an impurity solver,\cite{JAWhite} it  
has also been successfully used to combine 
dynamical mean field theory (DMFT) and {\em ab initio} 
techniques.\cite{flex1,flex2,flex3}
It is computationally relatively cheap, 
can be extended easily to more than two orbitals, and is also able 
to go beyond perturbation theory and give a more precise estimate for
the Kondo temperature. 

As we demonstrate, the performance of FLEX is good, and it is also 
able to capture the ST transition. However, while it 
automatically guarantees current conservation,
its convergence properties seem to be worse than those 
of simple iterated perturbation theory, and it is computationally also
more demanding. Nevertheless, in spite of these weaknesses, 
FLEX provides a very good option to study 
correlated behavior in nanoscale structures, 
and seems to provide a more accurate estimate for 
the Kondo temperature.

The paper is organized as follows. In Secs.~\ref{model} 
and \ref{flexsec},
we introduce the non-equilibrium two-level Anderson model, 
and describe the 
fluctuation exchange approximation  
used to solve the non-equilibrium Anderson model. 
In Sec.~\ref{iter} we show the details of the iteration of the full
Green's function within the  
fluctuation exchange approximation. 
In Sec.~\ref{sym}, we present the results obtained 
for completely symmetrical quantum dots with equal 
level widths, while in Sec.~\ref{assym}  
 results for dots with more generic parameters are discussed. 
Our conclusions are summarized in Sec.~\ref{conclu}, 
 and  some technical details are given in the Appendix.

\section{Theoretical framework}

\subsection{Model}\label{model}

Let us start by defining the Hamiltonian we 
use to describe the quantum dot. We divide the 
Hamiltonian into  a non-interacting part, $H_0$, and 
an interacting part, $H_{\rm int}$, and write $H_0$ as 
\be 
 H_{0}=H_{\rm cond}+H_{\rm hyb}+H_{0,\rm dot}\label{ham}\;. 
\ee
Here the term 
\begin{equation}
H_{0,\rm dot} = \sum_{i,\sigma}\varepsilon_{i}d_{i\sigma}^{\dagger}d_{i\sigma}\;,
\end{equation}
describes the individual levels of an isolated quantum dot, 
and correspondingly, $d_{i\sigma}^{\dagger}$ is the creation operator
  of a dot electron of spin $\sigma$ on level $i=\pm$, with energy 
$\varepsilon_{i}$. The other two terms, the conduction electron part, 
$H_{\rm cond}$, and the hybridization, $H_{\rm hyb}$, depend slightly 
on the geometry of the dot. For lateral dots, 
\begin{eqnarray}
   H^{\rm lat}_{\rm cond} & = & \sum_{\xi,\alpha,\sigma}
\xi_{\alpha}c_{\xi\alpha\sigma}^{\dagger}c_{\xi\alpha\sigma}\label{latcond}\;,\\
   H^{\rm lat}_{\rm hyb} & = & \sum_{\alpha,i,\xi,\sigma}t_{\alpha
     i}(c_{\xi\alpha\sigma}^{\dagger}d_{i\sigma}+h.c.)\;.
\end{eqnarray}
Here $\xi$ denotes the energy of a conduction electron measured from 
the (equilibrium) chemical potential of the leads, and
correspondingly, $c_{\xi\alpha\sigma}^{\dagger}$ creates a conduction
electron of spin $\sigma$ in lead $\alpha=L,R$.
In the presence of 
a bias voltage, this energy shifts to $\xi_\alpha = \xi + e V_\alpha$,
with $V_\alpha$ the electrical potential of lead
$\alpha$.\footnote{Notice, however, that the occupation continues to
  depend  on $\xi$, 
$\langle c_{\xi\alpha\sigma}^{\dagger}c_{\xi'\alpha\sigma}\rangle = 
\delta_{\xi,\xi'} \; f(\xi)$ with $f$ the Fermi function.}
The hybridization term $H^{\rm lat}_{\rm hyb}$  describes
tunneling between the dot level and the non-interacting leads, and the
parameters $t_{\alpha i}$ characterize the tunneling amplitude.

The terms $H_{\rm cond}$ and $H_{\rm hyb}$
are slightly different for vertical quantum dots or
carbon nanotubes. In these latter cases, each dot state is associated with a 
separate electron channel in each lead, $c_{\xi \alpha\sigma}\to c_{\xi i\alpha\sigma}$,
\begin{eqnarray}
   H^{\rm vert}_{\rm cond} & = & \sum_{\xi,i,\alpha,\sigma}\xi_{\alpha}c_{\xi i\alpha\sigma}^{\dagger}c_{\xi i\alpha\sigma}\label{vertcond}\;,\\
   H^{\rm vert}_{\rm hyb} & = & \sum_{\xi,i,\alpha,\sigma}
t_{\alpha i}(c_{\xi i\alpha\sigma}^{\dagger}d_{i\sigma}+h.c.)\;.
\end{eqnarray}

In this paper, we assume that the occupation of the two levels
involved in the transition is around $\langle \sum_{i,\sigma}
d^\dagger_{i\sigma}d_{i\sigma}\rangle \approx2$. Therefore, we write 
the interaction in an electron-hole symmetrical form\cite{prev}
\begin{equation}
 H_{\rm int} =
 \frac{U}{2}\left(\sum_{i\sigma}n_{i\sigma}-2\right)^{2}-J\;\vec{S}^{2}\;,
\label{intham}
\end{equation}
with $U$ and $J$ denoting the Hubbard interaction and the Hund's rule 
coupling, respectively, and 
$\vec{S}= \frac 12
\sum_{i,\sigma,\sigma'} d^\dagger_{i\sigma}\vec{\sigma}_{\sigma\sigma'} 
d_{i\sigma}$ being the spin of the dot.
 To carry out a systematic perturbation theory, 
we split the interaction above into a normal ordered term 
and a level shift,  
\begin{equation}
 H_{\rm int}= :H_{\rm int}: -\left(\frac {3U} 2 +    
\frac {3J} 4\right)  \sum_{i\sigma}n_{i\sigma} \;. 
\label{:intham:}
\end{equation}
We then incorporate the second term in $H_0$, 
\bea
H_0-\left(\frac {3U} 2 +    
\frac {3J} 4\right)  \sum_{i\sigma}n_{i\sigma}&\Rightarrow &\tilde H_0\;,
\\
\varepsilon_i -\left(\frac {3U} 2 +    
\frac {3J} 4\right)& \Rightarrow& 
\tilde \varepsilon_i\;, 
\eea
while we treat the normal ordered part
\begin{equation}
:{H}_{\rm int}:=\sum_{\scriptstyle i,j,m,n,\atop
\sigma,\sigma',\tilde\sigma, \tilde\sigma'}
\frac{1}{4}\;
\Gamma_{i\sigma\;n\tilde\sigma}^{j\sigma'\;m\tilde\sigma'}
%\tilde\Gamma_{i\sigma\;m\tilde\sigma'}^{j\sigma'\;n\tilde\sigma}
d_{j\sigma'}^{\dagger}d_{m\tilde\sigma'}^{\dagger}d_{n\tilde\sigma}d_{i\sigma}\; 
\label{eq:vert_ham}
\end{equation}
as a perturbation. Here the bare interaction vertices, 
$\Gamma_{i\sigma\;n\tilde\sigma}^{j\sigma'\;m\tilde\sigma'}$ can be
expressed in terms of $U$ and $J$, with the explicit expressions 
derived in Ref.~\onlinecite{prev}.
The above procedure must be contrasted to the one we
followed in Ref.~\onlinecite{prev}, where  the second term of
Eq.~\eqref{:intham:}
has been treated through the application of a counterterm procedure. 
This counterterm procedure becomes unnecessary in FLEX, which is formulated
in terms of the full (dressed) Green's functions.

\subsection{Out of equilibrium fluctuation exchange
  approximation}\label{flexsec}

To describe the spectral and transport properties of the dot, we 
use a Green's function method. We thereby consider the 
Keldysh Green's functions of the dot electrons,
\be 
G_{i\sigma\kappa}^{j\sigma'\kappa'}(t-t')
\equiv -i \;\langle {\rm T}_K
d_{j\sigma'\kappa'}(t)d^\dagger_{i\sigma\kappa}(t')\rangle \;,
\label{DefG}
\ee 
with $\langle\dots\rangle $ denoting the average with respect to the
stationary density matrix,  $ {\rm T}_K$ the time ordering along the
Keldysh contour, and $\kappa$ and $\kappa'=1,2$ the Keldysh indices,
labeling the upper and lower Keldysh contours. 
Throughout this paper 
we shall consider the simplest case, where the Hamiltonian is spin rotation 
invariant. In this case, the Green's function is spin diagonal,  
\be
G_{i\sigma\kappa}^{j\sigma'\kappa'}(t-t') = \delta_\sigma^{\sigma'}\;
G_{i\kappa}^{j\kappa'}(t-t')\,. 
\label{G_spindiag}
\ee

The non-interacting Green's functions, $g_{i\kappa}^{j\kappa'}$ are
associated with  $\tilde H_0$, and can be determined analytically
(see Appendix \ref{hybrg} for their explicit form). They are
related to the full Green's functions through the Dyson equation, 
\be
{\mathbf{G}}^{-1}(\omega) = {\mathbf{g}}^{-1}(\omega) 
-{\mathbf{\Sigma}}(\omega)\;, 
\label{Dyson}
\ee
where we used a matrix notation,  $m_{i\kappa}^{j\kappa'}\to 
\mathbf{m}$, and 
introduced the Keldysh self-energy, ${\mathbf{\Sigma}}$.

Just as in Ref.~\onlinecite{prev}, the knowledge of $G$ enables 
us to compute the current through the dot by  using the Meir-Wingreen
formula, 
\begin{eqnarray}
   I & = &\frac{ie}{h}\sum_{i,j}\int\limits_{-\infty}^{\infty} d\omega
\,\Big[({\Gamma}^{L}_{ij}
-{\Gamma}^{R}_{ij})(G^{<})^j_i(\omega)+
\label{curexp}\\
& + &
(f_{L}(\omega){\Gamma}_{ij}^{L}-f_{R}(\omega){\Gamma}_{ij}^{R})
%((G^R)^j_i(\omega)-(G^A)^j_i(\omega)) \Big]\;,
((G^>)^j_i(\omega)-(G^<)^j_i(\omega)) \Big]\;,
\nonumber
\end{eqnarray}
with the  lesser and greater Green's functions defined in
the usual way in terms of the Keldysh Green's function in  
Eq.~\eqref{G_spindiag}:
\bea
(G^{>})^j_i &=& G^{j2}_{i1}\;,
\\
(G^{<})^j_i &=& G^{j1}_{i2}\;.
\eea
The functions  $f_{\alpha}(\omega)=f(\omega-e V_{\alpha})$ in Eq.~\eqref{curexp}
denote  the shifted Fermi functions in lead $\alpha$, and the 
matrices $\Gamma_{ij}^\alpha$ describe the decay of the dot levels. 
They are defined as 
\begin{equation}
   (\Gamma_{ij}^{\alpha})_{\rm lat} =2\pi\; N_{\alpha}\;t_{\alpha i}t_{\alpha j}^{*}\;,
\end{equation}
for lateral quantum dots, while they read as 
\begin{equation}
   (\Gamma_{ij}^{\alpha})_{\rm vert} = \delta_{ij} \;2\pi \;
  N_{\alpha i}\;|t_{\alpha i}|^2\;,  
\end{equation}
for vertical dots, with $N_\alpha$ and $N_{\alpha i}$ 
standing for  the density of states in the leads. We remark 
that the factor $N_\alpha$ can be eliminated 
by incorporating it in the tunneling parameters, 
$t_{\alpha i}N_\alpha^{1/2}\to \tilde t_{\alpha i}$, and the fields
$c_{\xi\alpha\sigma }N_\alpha^{1/2}\to \psi_{\xi\alpha\sigma }$.

\begin{figure}[b]
\includegraphics[width=6.5cm,clip=true]{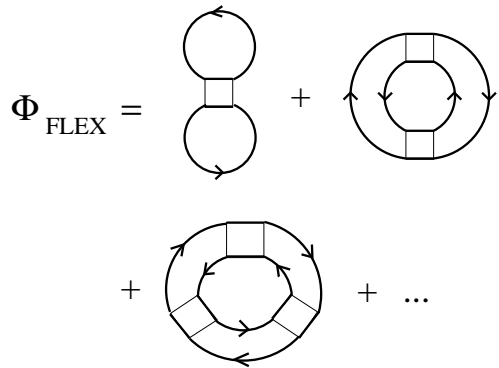}
\caption{The $\Phi$ functional generating the 
FLEX diagrams. The first diagram just generates 
the Hartree-Fock approximation. Heavy lines denote full Green's
functions. Squares denote the particle-hole vertex, defined 
in Eq.~\eqref{Gammatidle}.}\label{flex_generating}
\end{figure}
Our primary purpose is to determine $\mathbf{\Sigma}$
(and thus $\mathbf{G}$), and use that to compute the 
non-equilibrium differential conductance through the dot. 
We shall use the so-called  fluctuation exchange approximation 
(FLEX) for this purpose. 
FLEX is constructed in terms of 
a generating functional, $\Phi=\Phi[\mathbf{G}]$, defined as a functional of the
full many-body Green's function, $\mathbf{G}$.\cite{kb2} The self-energy and
the particle-hole irreducible vertex functions are obtained from
$\Phi$ through functional differentiation.  Although $\Phi$ is usually not
known, one can approximate it by a subset of diagrams, and then 
obtain approximations for the self-energy and the vertex functions.  
As shown by Kadanoff and Baym, \cite{kb1,kb2}
this construction is {\em conserving}, i.e., it 
guarantees that conservation laws are respected. Although 
this approach is mostly used in imaginary time, 
 one can quite naturally generalize it 
to the non-equilibrium case discussed here, by simply 
replacing the imaginary time 
Green's function in $\Phi$ by the Keldysh Green's functions.

\begin{figure}
\includegraphics[width=240pt,clip=true]{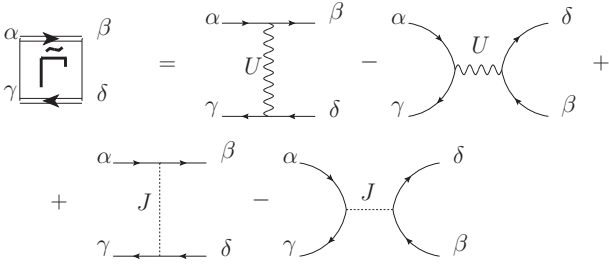}
\caption{Structure of the particle-hole  vertex, 
${\tilde \Gamma\;}_{\alpha \gamma}^{\beta\delta}$. Here 
$\alpha$, $\beta$, $\gamma$ and $\delta$ denote  composite 
indices as introduced in Eq.~\eqref{composite}}\label{phvertex}
\end{figure}

In this language, Hartree-Fock theory is just the simplest 
conserving approximation, while the next level of approximation 
is provided by FLEX, corresponding to the summation 
of an infinite series of ladder diagrams (see
Fig.~\ref{flex_generating}). 
In Fig.~\ref{flex_generating}. 
we introduced the Keldysh particle-hole vertex, 
\bea
\tilde{\Gamma\;}_{l_1,\sigma_1,\kappa_1\;l_2,\sigma_2,\kappa_2}^{l_3,\sigma_3,\kappa_3\;l_4,\sigma_4,\kappa_4}
&\equiv&s(\kappa_1)\;\delta_{\kappa_1\kappa_2\kappa_3\kappa_4}\;\tilde{\Gamma\;}_{l_1,\sigma_1\;l_2,\sigma_2}^{l_3,\sigma_3\;l_4,\sigma_4}\;, 
\nonumber
\\
\tilde{\Gamma\;}_{l_1,\sigma_1\;l_2,\sigma_2}^{l_3,\sigma_3\;l_4,\sigma_4} 
&\equiv&
 {\Gamma\;}_{l_1,\sigma_1\;l_4,\sigma_4}^{l_3,\sigma_3\;l_2,\sigma_2}\;,
\label{Gammatidle}
\eea
with $s(\kappa)$ keeping track of the sign change of the interaction 
on the Keldysh contour: $s(1)=+1$ for the upper and $s(2)=-1$ for the
lower contour. The structure of this particle-hole vertex, 
$\tilde \Gamma$, is shown in Fig.~\ref{phvertex} for the particular
case of Hund's rule coupling and  Hubbard interactions. 

\begin{figure}[b]
\centering\includegraphics[width=6.5cm,clip=true]{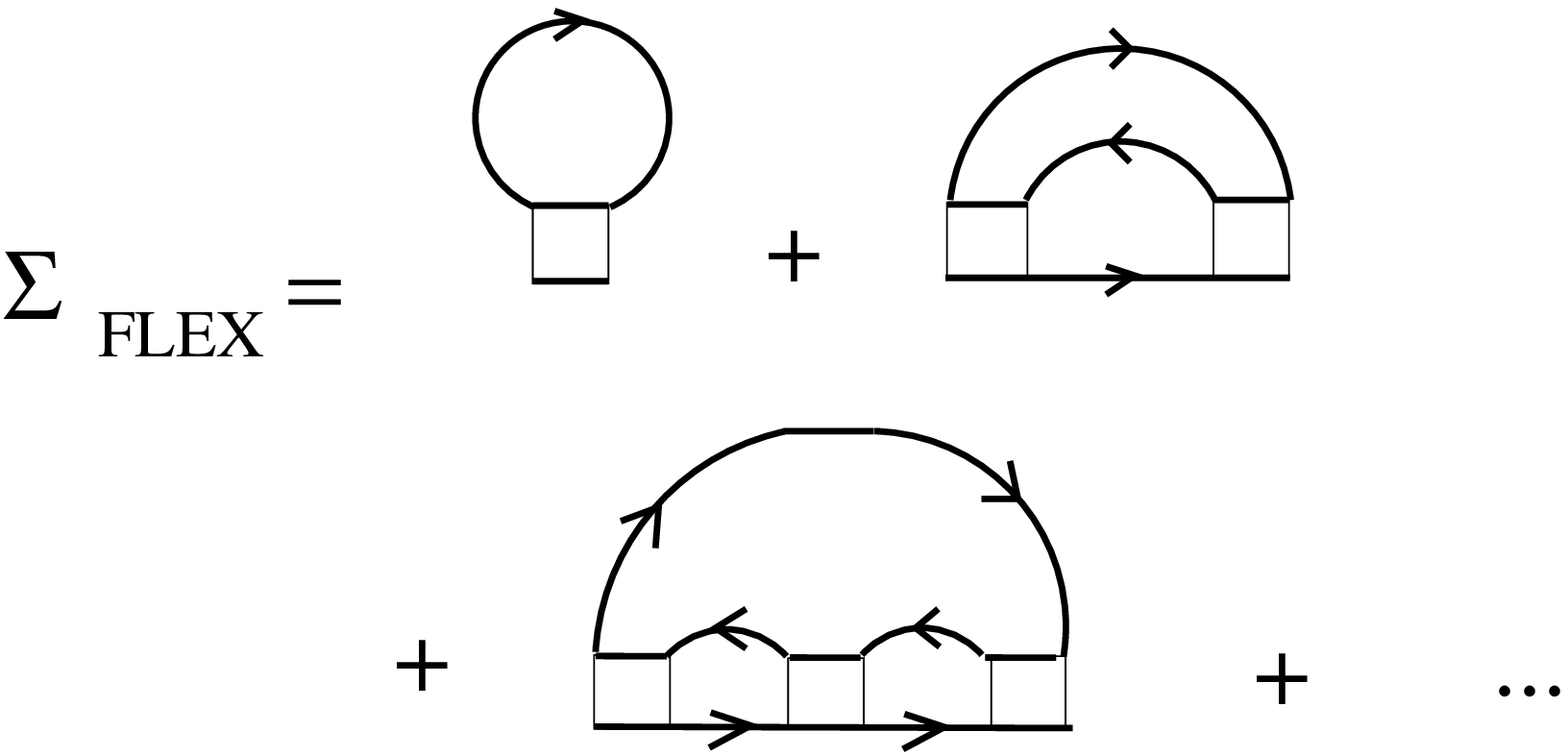}
\\
(a)
\null\vspace{0.3cm}
\\
\centering\includegraphics[width=6.5cm,clip=true]{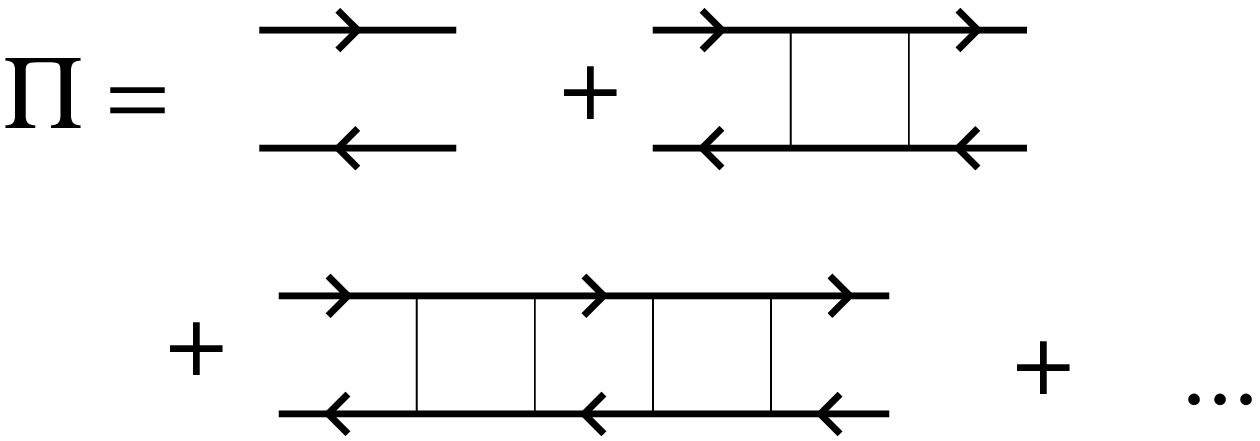}
\\
(b)
\caption{(a) Series of self-energy diagrams generated from  $\Phi_{\rm FLEX}$. 
(b) Diagrammatic definition of the full particle-hole propagator, 
$\Pi$. }
\label{self-energy}
\end{figure}
Differentiating the functional $\Phi$ of Fig.~\ref{flex_generating}, 
one obtains the  self-energy diagrams shown in Fig.~\ref{self-energy}.a. 
We then observe that all higher order diagrams contain  
the ladder series, shown in Fig.~\ref{self-energy}.b. 
Let us therefore introduce  the composite label,  
\be
(l_i,\sigma_i,\kappa_i)\to \alpha_i\;, 
\label{composite}
\ee
and define the particle-hole  propagator, $\Pi^{(0)}$ as 
\be 
{\Pi^{(0)}}_{\alpha_1\beta_1}^{\alpha_2\beta_2}(t-t')
\equiv i^{2}
G_{\alpha_1}^{\alpha_2}(t-t')G_{\beta_2}^{\beta_1}(t'-t)\;.
\label{polariz}
\ee
Then  the full particle-hole propagator, 
$\Pi$, defined by the ladder series in Fig.~\ref{self-energy}.b. 
satisfies the following Dyson equation: 
\bea
&& \Pi_{\alpha\beta}^{\alpha'\beta'}(t-t')
=  {\Pi^{(0)}}_{\alpha\beta}^{\alpha'\beta'}(t-t')
\label{Pi_Dyson}
\\
\nonumber
&& -i \sum_{\scriptstyle \alpha_1,\beta_1 \atop \alpha_2,\beta_2  }\;\int\limits_{-\infty}^{\infty} 
{\rm d}\tilde t\; {\Pi^{(0)}}_{\alpha\beta}^{\alpha_1\beta_1}(t-\tilde t)
\tilde{\Gamma}_{\alpha_1,\beta_1}^{\alpha_2,\beta_2}\;
{\Pi}_{\alpha_2\beta_2}^{\alpha'\beta'}(\tilde t-t')\;.
\eea
The integral being just a convolution, this 
equation can be solved in Fourier space. Defining then 
\bea
& & {\Sigma_{\alpha}^{\beta}}_{\rm ladder}(t-t')
\equiv 
\label{Sigma_Flex}
\\
&&\phantom{nn}-\sum_{\tilde \alpha, \tilde \beta}\sum_{\scriptstyle \alpha_1,\beta_1 \atop \alpha_2,\beta_2  }
\tilde{\Gamma}_{\alpha\tilde{\alpha}}^{{\alpha}_1 {\beta}_1}
{\Pi}_{\alpha_1{\beta}_1}^{\alpha_2\beta_2}(t-t')
\tilde{\Gamma}_{\alpha_2\beta_2}^{\beta\tilde{\beta}}
G_{\tilde{\alpha}}^{\tilde{\beta}}(t-t')\;,
\nonumber
\eea
 we can sum up all $n\ge 3$ order self-energy 
diagrams. The self-energy $\Sigma_{\rm ladder}$ also contains the second 
order self-energy contribution, but with double weight. 
Therefore, the total self-energy can be written as 
\be
 {\Sigma}  =  \Sigma_{\rm ladder}  +\Sigma^{(1)} - \Sigma^{(2)}\;, 
\label{Sigma_total}
\ee
with the first and second order diagrams,  $\Sigma^{(1)}$ and 
$\Sigma^{(2)}$ defined as 
\bea
& &{\Sigma^{(1)}}_{\alpha}^{\beta} 
      =i\;\delta_{\kappa_{\alpha}\kappa_{\beta}}s_{\kappa}\int\limits_{-\infty}^{\infty}
  \frac{d\omega_{1}}{2\pi}\sum_{\tilde{\alpha}\tilde{\beta}}{{\tilde{\Gamma}}}_{\alpha\beta}^{\tilde{\alpha}\tilde{\beta}}{G_{\tilde{\alpha}}^{\tilde{\beta}}}^{<}(\omega_{1})
\;,
\label{Sigma_1}\\
& & {{\Sigma^{(2)}}_{\alpha}^{\beta}}(t-t')\;
\label{Sigma_2}
\\
&=& -\frac 1 2
\sum_{\tilde \alpha, \tilde \beta}
\sum_{\scriptstyle \alpha_1,\beta_1 \atop \alpha_2,\beta_2  }
\tilde{\Gamma}_{\alpha\tilde{\alpha}}^{{\alpha}_1{\beta}_1}
{\Pi^{(0)}}_{\alpha_1{\beta}_1}^{\alpha_2\beta_2}(t-t')
\tilde{\Gamma}_{\alpha_2\beta_2}^{\beta\tilde{\beta}}
G_{\tilde{\alpha}}^{\tilde{\beta}}(t-t')\;.
\nonumber
\eea
 
Solving the equations above turns out to be numerically rather demanding 
for two reasons: First, to get a good enough time resolution, we have to keep 
a large number of time (frequency) points in the calculations. 
Second, the propagator $\Pi$ has too many indices. 
In fact, even in our simple case,    $\Pi$ has $8^4$ components. 
This number can be, however, substantially 
reduced if we exploit the SU(2) spin symmetry of the problem. 
Using simple group-theoretical 
arguments, we can show that the vertex  $\tilde \Gamma$ 
assumes a simple form  in spin space, and can be expressed in terms 
of a singlet and a triplet component,
\be
\ul{\ul{\tilde{\Gamma}}}_{\sigma_1\sigma_2}^{\sigma_3\sigma_4}=
\begin{bmatrix} \frac{1}{2}\left(\ul{\ul{\tilde{\Gamma}}}_{\;s}+\ul{\ul{\tilde{\Gamma}}}_{\;t}\right) 
 & 0 & 0 & \frac{1}{2}\left(\ul{\ul{\tilde{\Gamma}}}_{\;s}-\ul{\ul{\tilde{\Gamma}}}_{\;t}\right)\\
0 & \ul{\ul{\tilde{\Gamma}}}_{\;t} & 0 & 0 \\ 0 & 0 &
\ul{\ul{\tilde{\Gamma}}}_{\; t} & 0 \\
\frac{1}{2}\left(\ul{\ul{\tilde{\Gamma}}}_{\;s}-\ul{\ul{\tilde{\Gamma}}}_{\;t}\right) 
 & 0 & 0 & \frac{1}{2}\left(\ul{\ul{\tilde{\Gamma}}}_{s}+\ul{\ul{\tilde{\Gamma}}}_{t}\right)\;
\end{bmatrix}
\;,
\ee
with the four indices ordered as $\{\uparrow\uparrow, \uparrow\downarrow,\downarrow\uparrow,\downarrow\downarrow\}$, and the  
matrices $\ul{\ul{\tilde{\Gamma}}}_{s,t}$ 
defined as 
\bea
{\ul{\ul{\tilde{\Gamma}}}\;}_{t} & = & 
{\ul{\ul{\tilde{\Gamma}}}\;}_{\uparrow\uparrow}^{\uparrow\uparrow}-
{\ul{\ul{\tilde{\Gamma}}}\;}_{\downarrow\downarrow}^{\uparrow\uparrow}\;,
\\
{\ul{\ul{\tilde{\Gamma}}}\;}_{s} & = & 
{\ul{\ul{\tilde{\Gamma}}}\;}_{\uparrow\uparrow}^{\uparrow\uparrow}+
{\ul{\ul{\tilde{\Gamma}}}\;}_{\downarrow\downarrow}^{\uparrow\uparrow}\;. 
\eea
Here each entry is a matrix in the remaining orbital ($l$) and Keldysh
($\kappa$) labels: $(\tilde
\Gamma_{t,s})_{l_1\kappa_1;l_2\kappa_2}^{l_3\kappa_3;l_4\kappa_4}\to
{\ul{\ul{\tilde{\Gamma}}}\;}_{t,s}$.
By the same symmetry argument, we can show that 
 the propagators $\Pi^{(0)}$ and $\Pi$
take on a similar form. Furthermore, it is easy to see that 
this structure is maintained under multiplication, where the lower indices 
of a tensor are contracted with the upper indices of another tensor.
Therefore the
singlet and the triplet components of $\Pi$ can be summed up independently: 
\bea
 {\ul{\ul{\Pi}}\;}_{s,t}(\omega) 
  & = & {\ul{\ul{\Pi}}\;}_{s,t}^{(0)}(\omega)\left[\ul{\ul{1}}+i\;{\ul{\ul{\tilde{\Gamma}}}\;}_{s,t}\;{\ul{\ul{\Pi}}\;}_{s,t}^{(0)}(\omega)\right]^{-1}\;,
\label{Dyson_st}
\eea
with the unit matrix $\ul{\ul{1}}$ defined as 
$(\ul{\ul{1}})_{l_1\kappa_1;l_2\kappa_2}^{l_3\kappa_3;l_4\kappa_4} =
\delta^{l_3}_{l_1}\delta^{l_4}_{l_2}\delta^{\kappa_3}_{\kappa_1}\delta^{\kappa_4}_{\kappa_2}\;$.
 We can then simply express the  spin-independent 
part of $\Sigma_{\rm ladder}$ in terms of ${\ul{\ul{\Pi}}\;}_{s,t}$ as
\bea
&& ({\Sigma_{\rm ladder}})_{p}^{q}(t) = 
 -\sum_{\tilde{p}, \tilde{q}}
\frac{3}{2}\left(\ul{\ul{\tilde{\Gamma}}}_{\;t}\;\ul{\ul{\Pi}}_{\;t}(t)
\; \ul{\ul{\tilde{\Gamma}}}_{\;t}\right)_{p\tilde p }^{q \tilde q} 
{G\;}_{\tilde{p}}^{\tilde{q}}(t)
\nonumber
\\
 && \phantom{nnn}-\sum_{\tilde{p}, \tilde{q}}
\frac{1}{2}\left(\ul{\ul{\tilde{\Gamma}}}_{\;s}\;\ul{\ul{\Pi}}_{\;s}(t)
\; \ul{\ul{\tilde{\Gamma}}}_{\;s}\right)_{p\tilde p }^{q \tilde q}
\; {G\;}_{\tilde{p}}^{\tilde{q}}(t)\;,
\eea
with $p$ and $q$ denoting composite labels, 
only including the orbital and the Keldysh 
indices, $(l,\kappa)\to p,q$.

\subsection{Details of the FLEX iteration}
\label{iter}
The previously defined equations provide a self-consistent set of
equations, which we then solve iteratively. In 0'th order, we
approximate the full Green's function $\mathbf{G}$ by $\mathbf{g}$, 
\bea
 {G^{[0]}}_{\alpha}^{\beta}(\omega) & = &
 g_{\alpha}^{\beta}(\omega)\;,
\\
  {\Sigma^{[0]}}_{\alpha}^{\beta}(\omega) & = & 0\;.
\eea
We then start iteration $n\ge 1$, by first computing
$\mathbf{G}^{[n-1]}(t)$ from the Green's function $\mathbf{G}^{[n-1]}(\omega)$ 
of the previous iteration, by performing 
a Fast Fourier Transformation (FFT). Next, we construct 
$({{\ul{\ul\Pi}}}_{\;s,t}^{(0)})^{[n-1]}(t)$, obtain from that 
$({{\ul{\ul\Pi}}}_{\;s,t}^{(0)})^{[n-1]}(\omega)$,  and   
then we solve the Dyson equation, Eq.~\eqref{Dyson_st} to get 
${{\ul{\ul\Pi}}}_{\;s,t}^{[n-1]}(\omega)$.  From that we obtain 
${{\ul{\ul\Pi}}}_{\;s,t}^{[n-1]}(t)$ by FFT.  We can then use 
${\underline{\underline \Pi}}_{\;s,t}^{[n-1]}(t)$, 
$({{\ul{\ul\Pi}}}_{\;s,t}^{(0)})^{[n-1]}(t)$, and 
$\mathbf{G}^{[n-1]}(t)$ to compute 
$\mathbf{\Sigma}^{[n]}_{\rm ladder}$, $(\mathbf{\Sigma}^{(1)})^{[n]}$, and
$(\mathbf{\Sigma}^{(2)})^{[n]}$,  
and finally  the total self-energy, $\mathbf{\Sigma}^{[n]}(t)$,
through equations Eq.~\eqref{Sigma_Flex}, \eqref{Sigma_1}, 
\eqref{Sigma_2} and \eqref{Sigma_total}. 
Finally, we obtain our next estimate,
$\mathbf{G}^{[n]}(\omega)$, by first computing the Fourier transform, 
$\mathbf{\Sigma}^{[n]}(\omega)$, and inverting the Dyson equation, 
Eq.~\eqref{Dyson}. This iteration procedure is repeated until
convergence is reached.

    In the numerical calculations we represented the Green’s
functions using a finite uniform mesh of $N$ frequency points
in the range $-\Omega/ 2 < \omega < \Omega/ 2$. 
As mentioned above, the numerics was highly demanding; 
we had to use $2^{16}-2^{18}$ frequency points and $\Omega\approx1000U$ 
to reach convergence. 
The memory demand of the calculation was also much higher than that of 
the iterative perturbation theory (IPT) procedure of Ref.~\onlinecite{prev}. 
With the symmetry-based representation 
of $\tilde{\Gamma}$ and $\Pi$ propagators, however,  
we managed to reduce the size of them substantially and were able to
run the calculation on simple PC's.

Although for small interaction parameters the convergence was rather 
stable,  FLEX showed instabilities for high interaction parameters,
similar to IPT\cite{prev}.  These instabilities could be partially
cured by a gradual increase of the interaction parameters. 
With this trick, the range of applicability was found to be 
roughly the same as the one  found  with IPT.\cite{prev}

\section{Results and Discussion}\label{results}

\begin{figure}[b]
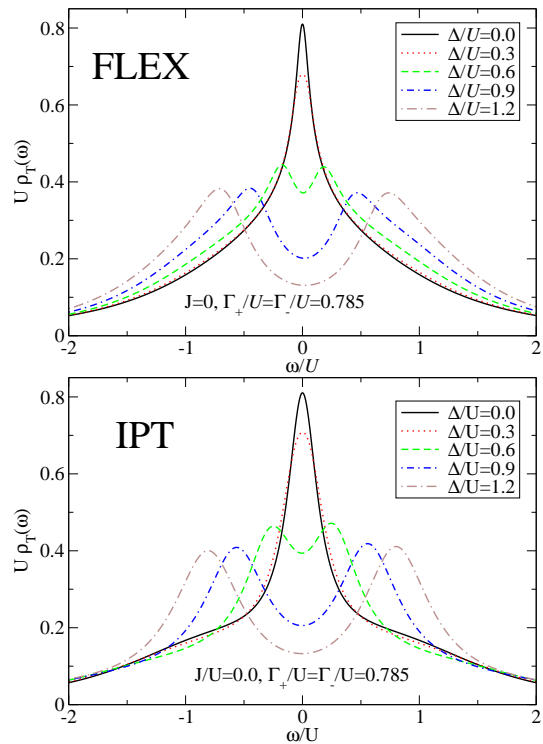

\includegraphics[width=200pt,clip=true]{fig5.eps}
\includegraphics[width=200pt,clip=true]{fig6.eps}
\caption{Total spectral function, $\rho_{T}(\omega)=(\rho_{+}(\omega)+\rho_{-}(\omega))/2$ for $J/U=0$ and
  $\Gamma_{\pm}/U=0.785$ for different values of level splitting,   
$\Delta/U$. On the top we show the FLEX results, while the bottom
shows the IPT results for the same parameters.}
\label{spect0}
\end{figure}

Let us now turn to the presentation of the numerical 
results. For simplicity, excepting Subsection~\ref{sec:asymmetrical}, 
in this section  we shall focus 
on a completely symmetrical dot with an even ($i=+$) and an odd 
($i=-$) level. In this case, the tunneling matrix elements 
satisfy 
\bea 
t_{L\pm} &=&\pm \; t_{R\pm}\;,
\eea
and the tunnelings can be characterized simply by the widths 
of the levels, 
\be 
\Gamma_i \equiv \sum_{\alpha=L,R} \Gamma^\alpha_{ii}\label{eq:Gamma_lat}\;,
\ee 
both for lateral and for vertical dots. 
Similar to
Ref.~\onlinecite{prev}, here we shall focus 
onto the vicinity of the 
electron-hole symmetrical point, 
$\tilde{\varepsilon}_+= \tilde{\varepsilon}_-= 0$, and assume that the  two 
levels are symmetrically positioned, 
\be 
\tilde{\varepsilon}_\pm = \pm \Delta/2\;.
\ee

\subsection{The case $\Gamma_{+}=\Gamma_{-}$}
\label{sym}

\subsubsection{Equilibrium spectral functions}

In this case, for $\Delta=J=0$, the three singlet and the triplet states 
of an isolated doubly occupied dot are completely degenerate, 
and an unusual Kondo state is  formed.\cite{prev,avishai} 
Turning on $\Delta$, one separates the
singlet state with both electrons on state $i=-$ from the rest
of the states, and destroys the Kondo effect once $\Delta$
becomes larger than the Kondo temperature, $T_K^*$, 
defined as  the halfwidth of
the central peak for $\Delta=J=0$. This transition can be
observed in the total equilibrium spectral functions,  
\begin{equation}
\rho_{T}(\omega)=-\sum_{i=\pm}\frac{1}{2\pi}\mathrm{Im}
\;G_{\;i,i}^{R}(\omega)\;,
\end{equation}
where the retarded Green's function is defined as, 
\be
G_{i,j}^R \equiv G^{i1}_{j1} -  G^{i1}_{j2} =   
{G}_{i,j}^{T} -  {G}_{i,j}^{<}\;,
\ee
with ${G}_{i,j}^{T} \equiv{G}^{i1}_{j1}$ the time-ordered Green's
function.

\begin{figure}[b]
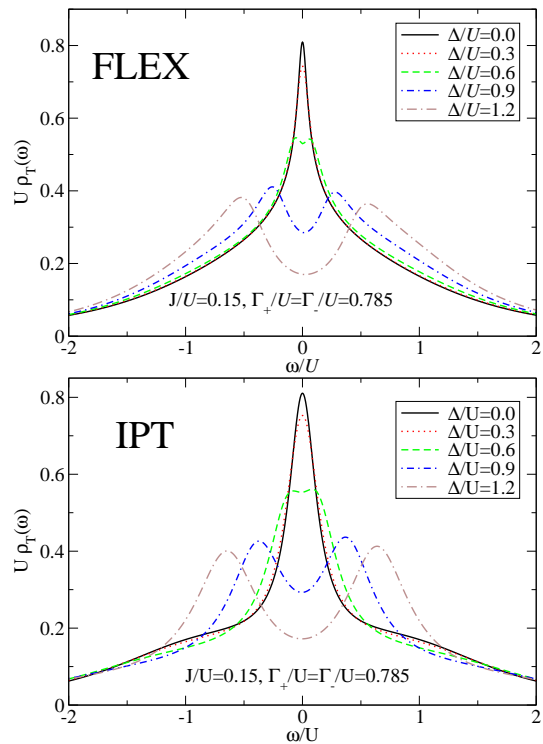

\includegraphics[width=200pt,clip=true]{fig7.eps}
\includegraphics[width=200pt,clip=true]{fig8.eps}
\caption{Total equilibrium spectral function, $\rho_{T}(\omega)$ for $J/U=0.15$ and $\Gamma_{\pm}/U=0.785$ for different values of level splitting, 
$\Delta/U$, as computed by FLEX (top) and by IPT (bottom).}
\label{spectj}
\end{figure}

In Fig.~\ref{spect0}, we display $\rho_{T}(\omega)$   
for  $J=0$ for various splittings of the two levels, 
$\Delta$, as computed by FLEX and by the 
iterated perturbation theory (IPT)  of Ref.~\onlinecite{prev}. 
The splitting of the Kondo resonance is remarkably similar in the two
figures, however, there are important differences, too. First of all, 
FLEX gives a smaller Kondo temperature, and provides a more  
realistic shape for the Kondo resonance both in the absence and in the presence of splitting. However, while the Hubbard peaks 
at $\omega=\pm U$ are still visible within the simple 
perturbative calculation, FLEX is unable to capture them correctly.

Similar conclusions are reached for $J\ne 0$ with the exception that
now the splitting of the Kondo resonance is shifted 
to higher values of $\Delta$ (see Fig.~\ref{spectj}). However, 
% as discussed in  Ref.~\onlinecite{prev}, 
in this case the central peak has a
slightly different interpretation than for $J=0$, since for 
$J>0$ the isolated dot would be in a triplet state. As a result, 
the central Kondo resonance at $\Delta=0$ can be interpreted as a
result of a triplet Kondo effect, where the spin $S=1$ of the 
dot is screened by the even and the odd conduction electron channels. 
In this triplet state the ground state degeneracy of the isolated dot
is reduced, 
and quantum fluctuations are therefore somewhat suppressed. 
As a consequence, the Kondo temperature $T_K$ is also 
reduced, and the central peak becomes slightly narrower, but   
also more stable against $\Delta\ne0$; in this $J>0$ case  the
splitting of the triplet Kondo resonance  occurs 
roughly  when $\Delta\sim 2 J + T_K$.

\subsubsection{Comparison with numerical renormalization group}

\begin{figure}
\includegraphics[width=6.5cm,clip=true]{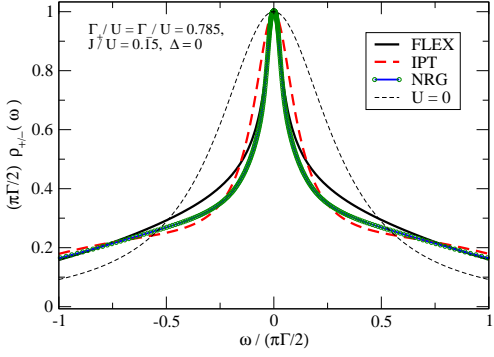}
\caption{Comparison of the spectral functions for $\Delta=0$, 
$\Gamma_\pm/U = 0.785$, and $J/U=0.15$, as computed by FLEX, IPT, and NRG. 
Clearly, FLEX seems to capture rather accurately the width 
of the central Kondo resonance.}\label{fig:NRG}
\end{figure}

Before entering the  discussion of 
the non-equilibrium results, it is worth 
comparing FLEX with 
other methods such as iterative perturbation theory (IPT), 
or numerical renormalization  group calculations (NRG),\cite{Wilson,Bulla} 
the latter procedure giving us a 
benchmark for the equilibrium calculations. 
Fig.~\ref{fig:NRG} compares
the results of these three methods for parameters
$\Delta=0$, $\Gamma_\pm/U = 0.785$, and $J/U=0.15$. 
For the NRG calculations we used the open access Budapest NRG 
code.\cite{BpNRG}  To reduce computational effort and 
achieve sufficient accuracy, we made use of the spin 
SU(2) symmetry of the Hamiltonian,  as well as 
the $U(1)$ symmetries corresponding to the 
conservation  of the total fermion numbers in channels $i=\pm$.
The computations were performed with a discretization parameter, 
$\Lambda=2$, and 2400 kept multiplets. The
calibration of the NRG parameters requires special care, since the 
NRG discretization and iteration procedure  renormalizes somewhat 
the bare parameters of the Hamiltonian.\cite{Wilson,Oliveira}
We  calibrated the level widths $\Gamma_\pm$
from the height of the numerically calculated spectral functions. 
The results obtained this way were in good agreement with the 
analytical expressions of Ref.~\onlinecite{Oliveira}. 

As shown in Fig.~\ref{fig:NRG} the width of the Kondo resonance
is perfectly captured by FLEX for the above parameters, while 
IPT slightly overestimates the size of the Kondo resonance. 
(As a comparison, in Fig.~\ref{fig:NRG} we also plotted the shape of 
the resonance for $U=0$.) However, while FLEX seems to give a better 
estimate for the Kondo temperature than IPT, IPT seems to capture the 
high-energy features (Hubbard peaks) better -- a well-known shortcoming
of FLEX.\cite{JAWhite}

\subsection{The asymmetric case, $\Gamma_{+}\neq\Gamma_{-}$}\label{assym}

\begin{figure}[b]
\includegraphics[width=240pt,clip=true]{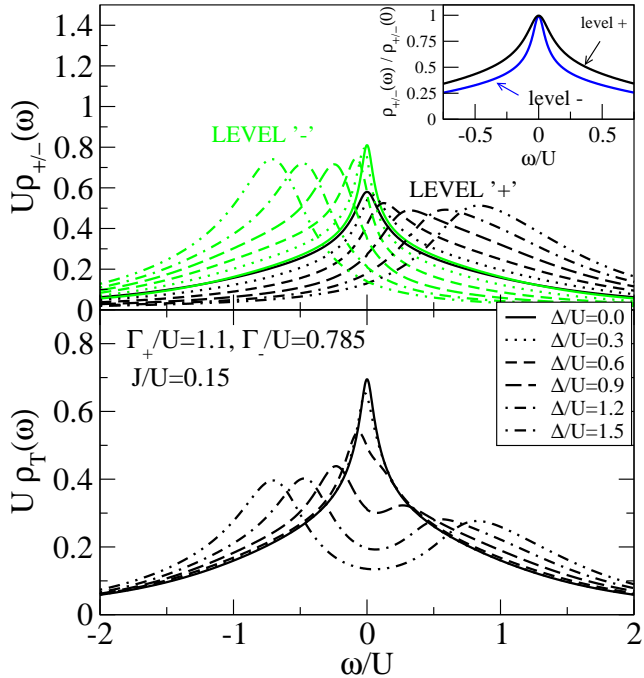}
\caption{Level-resolved (top) and total (bottom) equilibrium 
spectral functions  for $J/U=0.15$,  $\Gamma_{+}/U=1.1$ 
and $\Gamma_{-}/U=0.785$  for different level splittings, $\Delta/U$, as computed by FLEX. The inset in the upper panel shows the 
normalized spectral functions  $\rho_\pm(\omega)/\rho_\pm(0)$, 
demonstrating the presence of the two different Kondo scales. 
%\label{norm_spec}
\label{spect_as}}
\end{figure}

Let us now turn to the more generic situation,
$\Gamma_{+}\neq\Gamma_{-}$, and $J>0$. In this case, 
for $\Delta\approx 0$, the triplet spin on the dot is screened by a
two-stage Kondo effect,\cite{Pust} and the central resonances in the 
level-projected spectral functions, $\rho_\pm(\omega)$,
 become different due to the
presence of two different Kondo scales, $T_K^\pm$, corresponding 
to the screening in  the even and in the odd channels, 
respectively.  

\begin{figure}[b]
\includegraphics[width=240pt,clip=true]{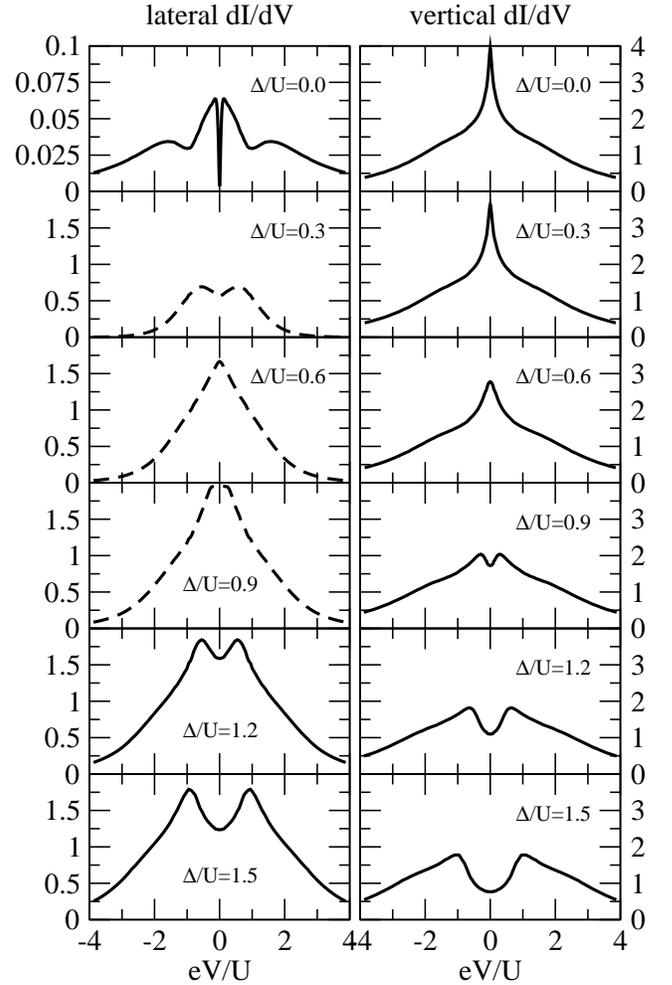}
\caption{Differential conductance, $G(V)={\rm d}I/{\rm d}V$,  
in units of $e^2/h$ for lateral (l.h.s.) and 
vertical (r.h.s.) dots, with $J/U=0.15$,
$\Gamma_{+}/U=0.785$, and $\Gamma_{-}/U=1.1$ for different level splittings, $\Delta/U$, as obtained by FLEX.
On the triplet side ($\Delta=0$), 
the second Kondo scale emerges as a narrow dip/sharp resonance in $G(V)$
in the lateral/vertical arrangement. For large $\Delta$'s the 
singlet-triplet splitting gives rise to a wide central dip in 
$G(V)$.  The curves reproduce very nicely all 
experimentally observed features, however, the cross-over
regime is only qualitatively captured in the lateral case.}
\label{vjassym}
\end{figure}

In Fig.~\ref{spect_as} we show the  level-projected
as well as  the full spectral functions, $\rho_T(\omega)$, 
as computed by FLEX for a dot with  $J/U=0.15$, 
 $\Gamma_{+}/U=1.1$ and $\Gamma_{-}/U=0.785$ 
for different level splittings, $\Delta/U$.
Unlike for $\Gamma_+=\Gamma_-$, 
for $\Delta=0$ the projected spectral functions of the two levels 
are different, $\rho_+(\omega) \ne
\rho_-(\omega)$. Nevertheless,  
they are all  symmetrical as a consequence
of a discrete particle-hole symmetry (see Ref.~\onlinecite{prev}). 
However, this symmetry is violated for any $\Delta\ne0$, where
electron-hole symmetry is destroyed even for the total spectral
function, $\rho_T(\omega)$. The difference in the Kondo
temperatures is clearly visible in the normalized level-projected 
spectral functions, 
shown in the inset of Fig.~\ref{spect_as}.

Similar to the symmetrical case, the Kondo resonances are gradually split
by a finite $\Delta$. The splitting of the resonances 
appears even more strikingly in the 
differential conductance, $G(V)={\rm d}I/{\rm d}V$, as computed from 
Eq.~\eqref{curexp}, and shown in Fig.~\ref{vjassym}. These differential 
conductance curves  were obtained by computing $\mathbf{G}$ and 
$I(V)$ for each bias voltage $V$ separately, and then
carrying out a numerical differentiation.

\begin{figure}
\includegraphics[width=6.5cm,clip=true]{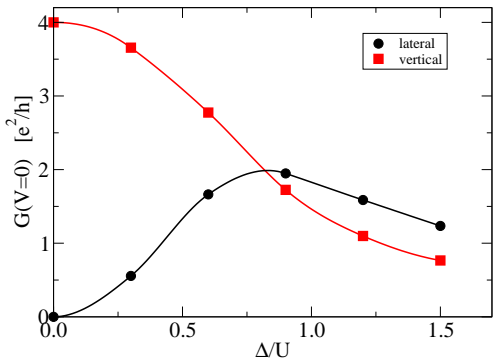}
\caption{
Linear  conductance, 
for lateral and 
vertical dots, with $J/U=0.15$,
$\Gamma_{+}/U=0.785$, and $\Gamma_{-}/U=1.1$ for 
different level splittings, $\Delta/U$, as obtained by FLEX.
\label{lincon}}
\end{figure}

For a lateral dot at $\Delta=0$, i.e. in the two-stage Kondo effect regime,  
the ${\rm d}I/{\rm d}V$ curve shows very nicely the build-up of the
first Kondo resonance,\cite{Wiel,Granger} and then the appearance
of a dip at  $V=0$ bias. This dip is a  result of the 
destructive interference between the two Kondo effects, and it appears
once the bias voltage becomes so  small that it cannot destroy even the 
narrower Kondo resonance of the spectral function. 
As shown in Fig.~\ref{lincon}, increasing $\Delta$, 
the linear conductance (i.e., the zero bias
differential conductance) exhibits  a maximum in the cross-over 
regime, in agreement
with the experiments. 
However, the bias-dependence of the differential
conductance in the cross-over regime (dashed lines in Fig.~\ref{vjassym}) of
maximal conductance is  not very reliable, and the $G(V)$ only shows
 the general  trends observed experimentally, i.e., the disappearance of 
the central dip, and the appearance of a state with a single Kondo 
resonance and 
a perfect $G=2e^2/h$ linear conductance. For even larger $\Delta$'s,
however, the ${\rm d}I/{\rm d}V$ curves show very nicely the linear splitting 
of the Kondo resonance.

In contrast to the lateral case, in a vertical geometry, 
the second Kondo effect manifests itself as an additional 
contribution to the conductance, and thus as a narrow peak at 
zero bias for $\Delta=0$. In this vertical case, the 
differential conductance curves reproduce the experimentally 
observed features even in the cross-over 
regime:  the linear conductance is suppressed with increasing $\Delta$ (see Fig.~\ref{lincon}),\footnote{for our parameters the two Kondo scales 
are very close to each other, and therefore this decrease is 
rather  featureless.}  
and  the central resonance gets gradually broader, until 
it splits into two side-peaks, corresponding to the singlet-triplet
excitation energy.

\begin{figure}
\includegraphics[width=6.5cm,clip=true]{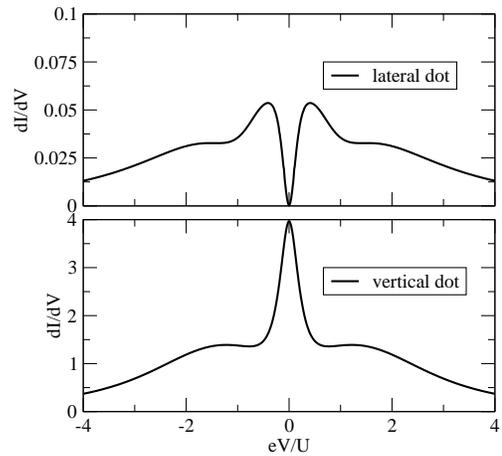}
\caption{Differential conductance (in units of $e^2/h$) 
of a vertical and a
lateral dot  with $J/U=0.15$,
$\Gamma_{+}/U=1.1$, $\Gamma_{-}/U=0.785$, and $\Delta=0$, as obtained
by IPT. The curves compare quite well with the ones in
Fig.~\ref{vjassym}.
}\label{didv_ipt}
\end{figure}

Finally, for a comparison, in Fig.~\ref{didv_ipt} we 
show the  ${\rm d}I/{\rm d}V$ curves at 
$\Delta=0$, as obtained by IPT, for the same parameters 
as the ones used to produce Fig.~\ref{vjassym}.
The IPT curves are strikingly 
similar in structure to the ones 
obtained by FLEX. The most important difference is
in the width of the central dip/resonance structure, 
which is somewhat narrower in 
the FLEX calculation, and is  closer to the real value.   

\subsection{The fully asymmetrical case}
\label{sec:asymmetrical}

So far, we focused on the case of a completely symmetrical quantum dot, 
and correspondingly, we assumed that one of the states is even while the other state is odd. In general, however, quantum dots are 
not entirely symmetrical. Such asymmetry leads to the suppression of 
the maximal conductance, and for lateral quantum dots it may also lead 
to interference effects.\cite{Meden}
It is out of the scope of the present paper to study such interference 
effects in detail, however, to demonstrate how FLEX works in 
this more general case, let us present here some results.

In this general case, we can parametrize the tunneling to the leads 
using the angles $\phi_\pm\in [-\pi/2,\pi/2]$ as 
\be
(t_{\pm,L}, t_{\pm,R}) = t_\pm \;(\cos(\phi_\pm),\sin(\phi_\pm))\;.
\ee
For an even level, $\phi=\pi/4$, while for an odd level, 
$\phi=-\pi/4$.  

\begin{figure}
\includegraphics[width=6.5cm,clip=true]{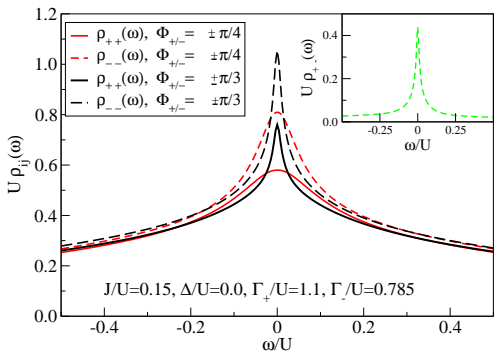}
\caption{Equilibrium dimensionless spectral functions for $J/U=0.15$
$\Gamma_{+}/U=1.1$, $\Gamma_{-}/U=0.785$, $\Delta/U=0$ 
for a symmetrical dot with $\phi_{\pm}=\pm\pi/4$
and for  an asymmetrical dot with  $\phi_{\pm}=\pm\pi/3$, as computed by FLEX.
The inset shows the off-diagonal component of the spectral function 
for $\phi_{\pm}=\pm\pi/3$.}\label{sp_compare}
\end{figure}

In Fig.~\ref{sp_compare} we present the equilibrium spectral 
functions, 
$$
\rho_{ij}(\omega)\equiv \frac{i}{2\pi}(G_{ij}^R(\omega)
-G_{ij}^A(\omega))\;,
$$ 
for the same level width, $\Gamma_+ /U= 1.1$, and 
$\Gamma_- /U= 0.785$ as before, but for a lateral dot with 
$\phi_\pm = \pm \pi/3$.
In this case left-right symmetry is absent, and $\rho_{ij}$
has offdiagonal components, too. 
Interference between the states 
$\pm$ appears as a resonant structure in $\rho_{+-}$. 
However, in contrast to the components $\rho_{++}$ and $\rho_{--}$,
within numerical accuracy 
$\rho_{+-}$
and $\rho_{-+}$ integrate to zero according to the 
corresponding spectral sum rule. 
For $\Delta=0$ the dot is still electron-hole symmetrical, 
and the heights of the spectral functions at $\omega=0$
are simply given by 
\be 
\rho_{ij}(0) = \frac {2}{\pi} (\mathbf{\Gamma}^{-1})_{ij}\;,
\ee
 with $\Gamma_{ij}= 
\sum_{\alpha=L,R}\Gamma^\alpha_{ij}$ the full relaxation rates 
(see Eq.~\eqref{eq:Gamma_lat}), as can be checked by an 
explicit calculation.

\begin{figure}
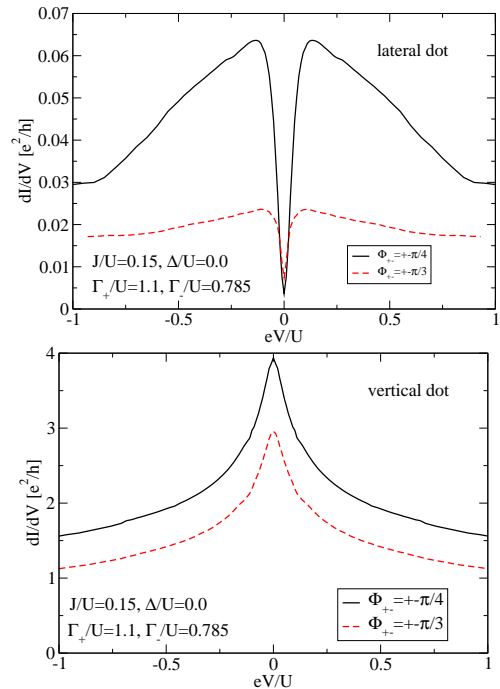

\includegraphics[width=6.5cm,clip=true]{fig15.eps}
\includegraphics[width=6.3cm,clip=true]{fig16.eps}
\caption{Top: Differential conductance, $G(V)={\rm d}I/{\rm d}V$,  
for lateral (top) and vertical (bottom) dots  with $J/U=0.15$,
$\Gamma_{+}/U=0.785$, $\Gamma_{-}/U=1.1$ and $\Delta/U=0$,
 as obtained by FLEX, for the symmetrical case, $\phi_{\pm}=\pm\pi/4$,
 and for the asymmetrical case, $\phi_{\pm}=\pm\pi/3$.
}
\label{didv_comp}
\end{figure}

Fig.~\ref{didv_comp} shows and compares 
the differential conductance computed for 
 asymmetric  vertical and lateral 
dots in the triplet regime ($\Delta=0$).
The curves are very similar to the ones obtained for symmetrical dots, 
excepting two important differences: (a) The conductance of a 
vertical dot does not reach the unitary conductance
but goes only up to the value $2e^2/h \;(\sin^2(2\phi_+) +\sin^2(2\phi_{-}))
= 3 e^2/h $, and similarly, the overall conductance of a lateral dot 
is also suppressed. (b) The width of the narrower resonance is 
reduced for a lateral dot. 
This is due to the fact that the smaller eigenvalues of the 
$\mathbf{\Gamma}$ matrix are reduced by the interference as 
\bea
\tilde \Gamma_{-}& =& \frac {\Gamma_{+} + \Gamma_{-}}{2}
\nonumber 
\\
&-&\sqrt{\frac {(\Gamma_{+} - \Gamma_{-})^2}{4} + \Gamma_{+}\Gamma_{-} \cos^2(\phi_+-\phi_-) }\;,
\nonumber 
\eea
and accordingly, the dip corresponding to the narrow Kondo 
resonance becomes also narrower. 
In contrast, the structure of the 
${\rm d}I/{\rm d}V$ curve remains essentially unaltered for a vertical 
dot, where only the amplitude of the signal is reduced.

\section{Conclusions}
\label{conclu}

In the present paper, we developed a general non-equilibrium
fluctuation exchange approximation (FLEX) formalism. 
We tested the performance of this approach on the 
singlet-triplet transition of a dot with two single-particle levels, 
driven by a competition between the Hund's 
rule coupling and the Kondo screening. This transition 
exhibits several correlation-induced features, which are typically 
rather difficult to capture. 
On the triplet side of the transition a Kondo state 
develops with two different Kondo scales,  while on the 
other side of the transition the triplet excitation appears as a
pseudogap feature. Finally, in the cross-over region an exotic  Kondo 
state appears, and for a lateral dot the linear conductance shows a 
broad resonance.

Remarkably, within its range of convergence, FLEX was able to capture 
all these features, excepting the Hubbard peaks, which are rather 
 poorly represented by FLEX. Nevertheless, the low energy features 
and the ${\rm d}I/{\rm d}V$ curves show behaviors remarkably close to 
the experimentally observed ones. In our earlier studies, we applied simple 
(iterative) perturbation theory (IPT) to describe the singlet-triplet 
transition. FLEX has some clear advantages, but also disadvantages 
with respect to IPT. On the one hand, it produces apparently more realistic 
curves in the small bias region than IPT, 
and -- as our comparison with NRG calculations confirms --
it captures the Kondo temperature as well as the Kondo effect-related structures better
there.
In addition, it is a generically conserving 
approximation, and it scales rather well with the number of orbitals. 
All these properties make FLEX a viable route to incorporate 
strong correlation effects in molecular electronics calculations. 
On the other hand, FLEX is computationally much 
more demanding. In fact, in this work we had to exploit symmetries 
to reduce the computational effort. 
This is, of course, not a major obstacle if one has access to 
supercomputers or efficient computer clusters, and we believe that the 
numerical efficiency can most likely be further improved.

Finally, let us comment on the version of FLEX we used here. 
In the present paper, we used a generating $\Phi$-functional, which only 
incorporates electron-hole bubble series. 
FLEX can, however, be extended to include fluctuations in
 the Cooper channel, too. 
This may be important in cases, where attractive 
interactions appear in some scattering channels. In particular, 
such an extension of FLEX may be necessary to describe 
transport through superconducting grains. The generalization is relatively straightforward, however, it is certainly  beyond the 
scope of the present work, which solely focused on 
 the demonstration of FLEX as an efficient non-equilibrium 
impurity solver.

\section{Acknowledgment}

This research has been  supported by the Hungarian Scientific 
Research Funds Nos.  K73361, CNK80991, NN76727,
T\'{A}MOP-4.2.1/B-09/1/KMR-2010-0002,  
and the EU-NKTH GEOMDISS project. G.Z. also acknowledges support 
from the Alexander von Humboldt Foundation. 
We would also like to thank Pascu Moca for kindly helping us to use special features of
the yet unpublished new version of the Budapest NRG code.

\appendix

\section{The hybridized Green's function, $\mathbf{g}$}\label{hybrg}
For completeness, let us give here the elements of $\mathbf{g}^{-1}(\omega)$. 
Restricting ourselves to the spin symmetrical case, 
$ {g^{-1}}_{i\sigma\kappa}^{j\sigma'\kappa'} = \delta_\sigma^{\sigma'} \;
{g^{-1}}_{i\kappa}^{j\kappa'}$.
 The elements of ${g^{-1}}_{i\kappa}^{j\kappa'}$ differ for lateral and 
vertical dots. For lateral dots, they are given by
\bea
{(g^{-1}_{\rm \;lat})}_{i\kappa}^{j\kappa'} &=&
{\delta\;}_i^js(\kappa) \left(\omega-\tilde \varepsilon_{i}\right)\delta_\kappa^{\kappa'} 
\nonumber \\
&- &
\sum_{\alpha\in L,R}N_\alpha \;t^*_{\alpha i}t_{\alpha j}\; \Delta_{\alpha}^{\kappa\kappa'}(\omega)\;,
\eea
with  $s(\kappa)$ the Keldysh sign defined in the 
main text, and  hybridization parameters 
$\Delta_{\alpha}^{\kappa\kappa'}(\omega)$ defined as 
\bea
 \Delta_{\alpha}^{11}(\omega) & = & \pi i(2f_{\alpha}(\omega)-1)\;,\\
 \Delta_{\alpha}^{12}(\omega) & = & -2\pi if_{\alpha}(\omega)\;,\\
 \Delta_{\alpha}^{21}(\omega) & = & -2\pi i(f_{\alpha}(\omega)-1)\;,\\
 \Delta_{\alpha}^{22}(\omega) & = & \pi i(2f_{\alpha}(\omega)-1)\;,
\eea
with  $f_{\alpha}(\omega)=f(\omega-e V_{\alpha})$ the shifted Fermi function.
For vertical dots, on the other hand,  ${g^{-1}}_{i\kappa}^{j\kappa'}$
is diagonal in $i$ and $j$, 
\bea
{(g^{-1}_{\rm \;vert})}_{i\kappa}^{j\kappa'} 
&=&
{\delta\;}_i^js(\kappa) \left(\omega-\tilde \varepsilon_{i}\right)\delta_\kappa^{\kappa'} 
\nonumber \\
&- &
{\delta\;}_i^j \sum_{\alpha\in L,R}N_{\alpha \;i} \;|t_{\alpha i}|^2\; \Delta_{\alpha}^{\kappa\kappa'}(\omega)\;.
\eea

\end{document}